# Luhmann's Communication-Theoretical Specification of the 'Genomena' of Husserl's Phenomenology

Loet Leydesdorff

Amsterdam School of Communications Research (ASCoR), University of Amsterdam, Kloveniersburgwal 48, 1012 CX  Amsterdam, The Netherlands.

loet@leydesdorff.net; http://www.leydesdorff.net

From the perspective of cultural studies and critical theory, Luhmann's communication-theoretical approach in sociology can still be read as a meta-biology: while biologists take the development of life as a given, Luhmann tends to treat the development of meaning as a cultural *given*.[1] Meaning is no longer considered as constructed in communication, but meaning processing precedes and controls communication as an independent variable. Habermas (1987) made the argument about this meta-biological foundation of Luhmann's systems theory most forcefully:

> In this way, subject-centered reason is replaced by systems rationality. As a result, the critique of reason carried out as a critique of metaphysics and a critique of power, which we have considered in these lectures, is deprived of its object. To the degree that systems theory does not merely make its specific disciplinary contribution with the system of the sciences but also penetrates the lifeworld with its claim to universality, it replaces metaphysical background convictions with metabiological ones. Hence, the

---

[1] Luhmann alternates between this meta-biological model using concepts of functional differentiation and structural coupling for the explanation, and a meta-theological one where meaning seems to be given transcendentally as a substance analogous to life (Luhmann, 1986). The meta-theological metaphor is pursued by Luhmann by grounding his theory on the operation of the distinction (that is, on a paradox). The *first* distinction is then Lucifer's breaking away from God as the devil (Luhmann, 1990, at pp. 118 ff.) or, in other words, the problem of the Theodicy, that is, the origin of evil in the world (Leibniz, [1710] 1962).

> conflict between objectivists and subjectivists loses its point. (Habermas, 1987, at p. 385).

In other words, Habermas appreciates Luhmann's distinction between psychic and social systems, but he challenges us to bring the critique of metaphysical issues (of providing meaning to events in a dialectics) back into this metabiological perspective that processes meaning without intentionality, that is, as a scientistic objectivation.

How can one think both these metaphysical and metabiological perspectives? I submit that the mathematical perspective can help bridge the gap between the meta-biological and the meta-physical perspectives without resolving the tension between the two. An additional degree of freedom can be specified and then used for the reflection. This reflection is different from the substantive reflections, since it is formal, that is, yet without meaning, and therefore cannot substantively support the construction of a meta-level. The additional axis which is used for the reflection is only spanned at ninety degrees and not as a mirror at 180 degrees. Consequently, this formal reflection is not 'meta'—that is, at a next level—but it remains 'epi': the added dimension opens a space of alternative possibilities for recombining the substantive dynamics in the previous communications of meaning. The historical operation of the system remains substantive.

I shall turn to Husserl's (1929) *Cartesian Meditations* for epistemological specification and foundation of this reflection in *inter*subjectivity. The *Cogito* of Descartes and Shannon's mathematization of uncertainty provide us with tools to understand the uncertainty in communication as a *cogitatum*, that is, a subject of doubt. The mathematization of anticipation (Dubois, 1998; Rosen, 1985) enables us to relate the uncertainty at each moment of time with expectation and anticipation over time without having to resort to a single—that is, undifferentiated—transcendentality. A system of social relations can be specified as a *cogitatum* which transcends individual minds but remains structurally coupled to them as their distribution. An individual mind cannot specify this system without entertaining a discourse. However, its possibility can fruitfully be hypothesized. A program of empirical studies and simulations can thus be formulated.

**Anticipation, uncertainty, and the intentionality of communication**

Both Luhmann and Habermas make references to Husserl's reflections on intersubjectivity (Husserl, 1929, 1936, 1962; Derrida, 1974). It has been claimed in the literature (Schutz, 1952, at p. 105; cf. Habermas, 1981, at p. 178f.; Luhmann, 1995, at p. 170) that Husserl failed to ground the concept of 'intersubjectivity,' but I follow Luhmann's (1995) interpretation that this was never Husserl's argument or intention.

The *locus classicus* for the alleged failure of Husserl is Alfred Schutz's (1952) study entitled 'Das Problem der transzendentalen Intersubjektivität bei Husserl' (The problem of intersubjectivity with Husserl; Schutz, 1975). Schutz formulated:

> All communication, whether by so-called expressive movements, deictic gestures, or the use of visual or acoustic signs, already presupposes an external event in that common surrounding world which, according to Husserl, is not constituted except by communication. (Schutz, 1975, at p. 72).[2]

Schutz wished to ground the communication in the 'life-world' as a common frame of reference for the communication. He criticized Husserl for explaining this ground as a *result* of communication. However, Husserl considered the external referent of communication as a horizon of a potential variety of meanings. Husserl's intersubjectivity remains intentional, whereas Schutz argued in favor of an existential grounding of intersubjectivity in a 'we,' for example, when he went on to say: 'As long as man is born from woman, intersubjectivity and the we-relationship will be the foundation for all other categories of human existence.' (*ibid*., at p. 82).[3]

In other words, Schutz disagreed with Husserl about the possibility of *deriving* social relations from communication. Social relations, in Schutz's opinion, are prior to communications, while Husserl argued that social relations are embedded in communication or—as he put it—'transcendental intersubjectivity.' Husserl abstracted from the historical given (that is, the modeled system) by putting the latter between brackets in what he called an ἐποχη (epoché). What is left when one abstracts from what is given? These non-material traces can be provided with meaning as the subject of a meditation. The meditation gives access to an *a priori* domain for which no discursive concepts are available to the Ego. In this meditation Husserl not only questions what it means to be 'human,' but also asks about the referent of human intentionality. While the first question refers back to Descartes' '*cogito ergo sum*,' the latter addresses the subject of doubt in the *cogito*, that is, the *cogitata*. In the *Cartesian Meditations* of 1929, this quest is elaborated by extending on Descartes' (1637) category of the *cogito*. In addition to oneself as the subject of doubt, one can also raise the question of the substance about which one is in doubt, that is, the *cogitatum*: the external referent or the subject of one's doubting.

---

[2] 'Alle Kommunikation, ob es sich um eine sogenannte Ausdrucksbewegung, eine Zeigegeste, oder den Gebrauch visueller oder akustischer Zeichen handelt, setzt bereits einen äußeren Vorgang in eben jener gemeinsamen Umwelt voraus, die nach Husserl erst durch die Kommunikation konstituiert werden soll.' (Schütz, 1952, at p. 97).
[3] 'Solange Menschen von Müttern geboren warden, fundiert Intersubjektivität und Wirbeziehung alle anderen Kategorien des Menschseins.' (Schütz, 1952, at p. 105).

For Descartes this *cogitatum* could be distinguished only negatively from the *cogito* as that which transcends the contingency of one's *cogito* at each moment of time. From this perspective, the Other in the act of doubting is defined as God. God transcends the contingency of the *cogito*, and therefore one can expect this Other to be eternal. While the *cogito* knows itself to be mortal and incomplete, God as the Other is necessarily complete. Because of His completeness, the world as His creation is given to us transcendentally. According to Descartes, it would contradict the completeness of God (not to mention His veracity) if the transcendental dimension of the world were a continuous deception. This *ex ante* correspondence between our ideas and experiental uncertainties was elaborated by Leibniz (1695) into the idea of an *harmonie préétablie*. Leibniz ($^3$1966, at p. 272) acknowledged that this idea has the status of a hypothesis among other possible arrangements, but entertaining this hypothesis was the only option for explaining the perfection of Nature as God's creation (Leydesdorff, 1994, at p. 36). Kant agreed with this superior likelihood of Leibniz's explanation above others, and designated it as the cosmopolitan proof of the existence of God. (Descartes's position was characterized by Kant as the ontological proof of the existence of God. Both proofs, however, were rejected by Kant as insufficient evidence for the existence of God.)[4]

Husserl proposed in the *Meditations* to consider the *cogitatum* as a self-reinforcing ground for the meditation. It provides the *cogito* with a reference to a horizon of meanings (unlike the single Transcendency of Descartes). We are uncertain about what things mean. and this generates an intersubjectivity which transcends individual subjectivity. Although meanings are structured at the supra-individual level, these structures are no longer necessarily identified with a personal God.[5] On the contrary, meaning can be constructed, enriched, and reproduced among human beings by using language. Husserl acknowledged this function of language for the generation of meaning when he formulated for example: 'The beginning is the pure and one might say still mute experience which first has to be brought into the articulation of its meaning' (*ibid.*, p. 40).

Unlike Habermas who proposed to ground (Schutz's) 'lifeworld' in language, Husserl grounds the act of doubt in *noesis*, that is, a knowledgeable awareness of the structuredness of experiences. This formal awareness generates an intentionality in the present which can be communicated among human beings as meaningful. Husserl calls this intentionality transcendental because it precedes its articulation in language. Language can then be considered as a first-order ('natural') medium for the communication. However, the *noesis* abstracts from everyday language by adding the reflexivity of the knowing subject who can remain aware of his/her intentionality in communicating. Note that this awareness accompanies the expression in communication at a level which remains 'epi' rather than 'meta.'

---

[4] In the terminology of the 17[th] century, the issue was formulated in terms of how the substances communicate. Descartes distinguished between two substances: matter (*res extensa*) and thought (*res cogitans*). In the Catholic religion the two are mediated in the miracle of transubstantiation. But how can the mind know about the world using scientific reasoning about uncertainties? Leibniz (1695) assumed an *a priori* correspondence originating from God's creation. Others (e.g., Huygens) argued that one has to distinguish analytically between mathematical clarity and experimental uncertainty. Hypotheses then can be tested (Huygens [1690], 1888-1950, Vol. XXI, at p. 541; Elzinga, 1972, at p. 37). From our sociological perspective, the question remains of what generates the hypotheses? Are these individual ideas, transpersonal mathematics, or intersubjective discourses? For Leibniz these three possible sources of variation were harmonized at a single (transcendental) origin (Leydesdorff, 1994). The evolutionary dynamics among these three dimensions were addressed from philosophical perspectives by Popper (1972) and Hull (2001), among others.
[5] I leave the relation with Spinoza's philosophy out of consideration here. For Spinoza, all structure is transcendental and therefore within God. This leads to pantheism.

By using language one is able to relate meanings to one another, but within language the world is resurrected as an architecture in which the words additionally can be provided with a position as concepts. This *cogitatum* can only be grasped by a reflexive agent in an intentional performance analogous to capturing the operation of one's own *cogito*. Husserl emphasized that the substance of the social system ('intersubjectivity') is different because one knows it *ex ante* as transcending the domain of the individual. The study of this new domain would provide us with 'a concrete ontology and a theory of science' (*ibid*., at p. 159). However, Husserl concedes that he has no instruments beyond the transcendental apperception of this domain and therefore he has to refrain from empirical investigation:

> We must forgo a more precise investigation of the layer of meaning which provides the human world and culture, as such, with a specific meaning and therewith provides this world with specifically "mental" predicates. (Husserl, 1929, at p. 138; my translation).

Intersubjectivity precedes objectivity in the world (*ibid*., at p. 160) because the world is represented within it, for example, by using language. Phenomena remain an instantiation. First, the experience of the phenomenological world may be common sense (for example, using natural languages), but the meanings provided to what is represented can further be codified as in scientific discourse. Thus, the system is grounded in inter-subjective knowledge that is generated historically within the system.

The order of priority itself changes with Husserl's reflection: the *cogitatum* provides a necessary condition for the *cogito*, although the latter remains a historical condition for the former. Husserl used the word 'rooting' for the historical origins, but he emphasized that the intentionality develops in the present. As noted, Husserl was not able to specify the 'mental' predicates which he hypothesized other than as an analogy to the categories of philosophical reflection within the *cogito*. However, the mathematical theory of communication provides us with the categories for studying communications in terms of uncertainties, and the theory of anticipatory systems provides us with categories for studying their evolution.[6]

First, Shannon's (1948) theory of communication can be elaborated for systems which communicate in more than a single dimension at the same time (Leydesdorff, 1995; Theil, 1972). Human language can be considered as the evolutionary step that enables us to communicate both (Shannon-type) information and meaning. Meaningful information can be distinguished from Shannon-type information as an interaction term between these two layers of processing (Leydesdorff, 2003). However, meaning is provided from the perspective of hindsight. Thus, the arrow of time is locally inverted. This can be modeled using Rosen's (1985) theory of anticipatory systems. Although Rosen's model was developed within theoretical biology, the model can be made more general by mathematization in terms of incursive routines (Dubois, 1998; Leydesdorff, 2005). The incursive operation is recursive: some meanings can be provided with meaning in a scientific discourse and thus be further codified.

Two anticipatory mechanisms at a systems level can co-evolve hyper-incursively into a strongly anticipatory system (Leydesdorff, forthcoming). The one anticipation is contained, for example, in the demarcation of scientific communications from differently coded ones (e.g., on the market), and the other precedes differentiation and is inherent to specifying meaning among human beings using language. Thus, the formal and informal layers of communication are always intertwined. However, they can be distinguished analytically. The analytical dimensions abstract from the observable world. One might say that the models provide us with a 'genomenology' in which we are able to specify the 'genotypes' as analytically different routines without a one-to-one relation with observable phenomena. The models remain our constructs in language and *noesis*.

---

[6] Künzler (1987, at p. 331) formulated that 'between Luhmann's marginalization of language (cf. Habermas, 1985, at p. 438) and Habermas's foundation of sociology in the theory of language, one should be able to find the comparatively innocent consideration of meaning as the *ratio essendi* of language and language as the *ratio cognoscendi* of meaning' (my translation). My argument, however, goes beyond this position because I argue that codified knowledge in a functionally differentiated configuration can only be analyzed by invoking codes of communication generalized at a level beyond human language.

**Husserl's phenomenology and the social genomenologies**

The formal perspective of mathematics which abstracts from the contents of respective models enables us to think of inter-subjectivity as an interaction effect grounded in an inter-textuality or even an inter-disciplinarity among perspectives. Phenomena can be expected to contain interaction terms among the genomena. The subdynamics of the genomenological organization of the complex system can first be abstracted analytically from the substantive categories that have been made available to us in the positive sciences (e.g., variation and selection) by appreciating these substantive insights as the specifications of subdynamics. The historical phenomena can then be reconstructed as the results of interactions among the reflections. Culture and civilization thus remain constructs that feed back on what is 'naturally' or previously given. The feedbacks operate in an anticipatory mode. The sciences are part and parcel of the knowledge bases of this transformative culture. The external references of the communication can be considered as a reality 'out there,' but these 'realities' have been constructed reflexively and therefore invested with meaning.

In economics, for example, this new meaning provided by the reconstruction can be appreciated as the value of commodities on the market. The values are shaped by market forces (under the conditions of modernity). Thus, in addition to commodities, capital and shares can also be traded. Since Newton the concepts of physics (like centers of gravity and gravitation) have been theoretical constructs that can be provided with an interpretation by the experiment. These concepts have highly codified meanings. Maturana (2000, at p. 169) formulated this as an 'interobjectivity that we observers in language live as operations in the flow of languaging in which we observers exist.' Our frame of reference, in my opinion, is not the 'we' of the aggregate of living subjects, but the interactive development of this 'interobjectivity' as a *cogitatum*.

The phenomena in this 'interobjectivity' can be considered as stabilizations based on quasi-equilibria among the fluxes of communications. The stabilizations occur historically as instantiations among other potentially possible ones. How can the knowledge-based operations which reproduce the phenomena as a knowledge-based system be specified? Let me quote Husserl one more time:

> This, the *cogitatum as cogitatum* can never be imagined as a readily available given; it becomes clear by the specification of the horizon and the continuously regenerated horizons. The specification itself is yet always imperfect, but because of its uncertainty to be considered as a structure of determination. For example, the cube leaves perspectives open on the sides which cannot be seen, and nevertheless as a cube conceptualized […]. This openness precedes the real specification in more detail. […] Real observing—unlike abstract clarification in the anticipating imagination—leads to more precise specification and perhaps differentiation, but with new horizons opening. (*ibid*., at pp. 47f.; my translation).

Note the emphasis on 'anticipating imagination.' The positive sciences (e.g., biology) tend to begin with observations, but in sociology this approach has led to positivism. On the basis of *a priori* decisions, the stabilization of facts would then be given priority over the meanings of these facts. This generates only one perspective among other possible ones. When this configuration is reflected, observations in the past can be turned into expectations about the future provided that a code in the communication is stabilized for making the inference.

Sociological specification is neither able nor allowed to forget that the 'facts' contain meanings and thus already imply a perspective. This double perspective is the very subject of methodological reflection in the discipline. Next-order, symbolic, and global horizons can be expected to resonate within hitherto stabilized meanings. Further reflections bring the meanings and the tensions among them to the fore. Meanings can be subjective and/or intersubjective. Intersubjective meanings can further be codified, for example, by using symbolically generalized media of communication in addition to everyday languages. For example, science can be considered in a sociology of science as the specific subsystem of communication in which truth-finding operates as a symbolically generalized medium of communication.

The meaning of intersubjectivity could be specified by Husserl only from the perspective of the subjective *cogito*, but nevertheless as a genomenon—a monade—that transcends the individual *cogito*. Other authors, e.g. Levinas (1953, 1963), have argued in favor of the foundation of intersubjectivity in the relation with the other human being ('Thou') as a transcendental instance. In sociology, however, the contingent other provides an instance of a double contingency, while the configuration can modulate this relation as a third source of the uncertainty. Three sources of uncertainty can no longer be envisaged within a single metaphor. Since the metaphors are then no longer 'natural,' the systems under study tend to become visible as knowledge-based constructions that remain under reconstruction.

In summary, the knowledge-based system remains the result of the intentionality of the carriers, but this is not a sufficient condition for the emergence of an order of expectations at the systems level. The groupings and the interactions of weak anticipations in social formations need first to be developed into a differentiated communication structure that contains another asynchronicity and, therefore, potential incursion endogenously. Insofar as this second incursion resonates with our weak anticipations and predictions, the intersubjectivity can become an interobjectivity. This remains always a matter of degree because the strongly anticipatory system *co*-constructs our future on the basis of organizational formats that are entertained in knowledge-based representations.

Because of the expectation of incompleteness and fragmentation of the codifications in the differentiated system, the formal reflection cannot guide us as a grandiose meta-theory. However, it is available and needed at the *epi*-level (that is, around) to increase the clarity of the substantive reflections. Simulations inform expectations in quasi-experiments in measuring the knowledge bases of historically stabilized systems. The models no longer need to be 'history friendly' or grounded in the evolutionary metaphor of survival. The emerging knowledge base is not a living system, but a social system. It rests like a hyper-network on the networks on which it builds. As a hypercycle, the knowledge-based subdynamics transform 'the ship while a storm is raging on the open sea.' In the resulting economy this reconstruction is no longer a mission only for the scientists involved (Leydesdorff, 2006). The matching of knowledge-based anticipations and innovative reconstruction on this basis has been woven into the complex dynamics of the social system and its economy as one of its coordination mechanisms.

**Conclusions**

I argued that Husserl's (1929) specification of intersubjectivity as a phenomenological *cogitatum* raises the question about its explanation. The *cogitatum as cogitatum* is not readily given, but it can be specified as an expectation. Shannon's (1948) mathematical definition of communication in terms of the expected information content of distributions and Rosen's (1985) definition of an anticipatory system as a system which entertains a model of itself, open the social system as a domain for investigation containing self-referential communications and potentially self-organizing control mechanisms. Luhmann (1984) formulated substantive hypotheses regarding this domain such as its potential to self-organize the communication when the symbolically generalized media of communication are functionally differentiated.

The non-linear dynamics of meaning-processing in social communication are complex since composed of several sub-dynamics. In addition to the functional differentiation, Luhmann (1997) specified a differentiation among interaction, organization, and self-organization. Organization can then be considered as the retention mechanism of the self-organization of communication. Without organization the uncertainty would no longer be historically controllable, and therefore the system would no longer be able to build evolutionarily on previous stages.

Stabilization and globalization of meaning-processing (over time) can be expected as genomena of communication in addition to variation and selection. Both layers—information-processing and meaning-processing—can be expected to operate with this full set of four sub-dynamics: variation, selection, stabilization, and globalization (Leydesdorff, 2001). Unlike information-processing, however, meaning is provided from the perspective of hindsight and thus against the axis of time. The production of probabilistic entropy within meaning-processing can therefore be negative. This remains an empirical question (e.g., Leydesdorff *et al.*, 2006).

Furthermore, the algorithms for the study of anticipatory systems in terms of incursive and hyperincursive routines, enable us increasingly to specify expectations about the non-linear dynamics of expectations (Dubois, 1998, 2000; Leydesdorff & Dubois, 2004; Leydesdorff, 2005 and forthcoming). For the first time in history, it seems that we thus have sophisticated theory and advanced methods for examining the substance of social communication, that is, the processing of meaning.

return

**References**

Derrida, J. (1974). *Edmund Husserl's origine de la géometrie*. Paris: Presses Universitaires de France.

Descartes, R. (1637). *Discours de la méthode*: Amsterdam.

Dubois, D. M. (1998). Computing Anticipatory Systems with Incursion and Hyperincursion. In D. M. Dubois (Ed.), *Computing Anticipatory Systems, CASYS—First International Conference* (Vol. 437, pp. 3-29). Woodbury, NY: American Institute of Physics.

Dubois, D. M. (2000). Review of Incursive, Hyperincursive and Anticipatory Systems—Foundation of Anticipation in Electromagnetism. In D. M. Dubois (Ed.), *Computing Anticipatory Systems Casys'99* (Vol. 517, pp. 3-30). Liege: Amercian Institute of Physics.

Elzinga, A. (1972). *On a Research Program in Early Modern Physics*. Gothenburg: Akademiförlaget.

Habermas, J. (1981). *Theorie des kommunikativen Handelns*. Frankfurt a.M.: Suhrkamp.

Habermas, J. (1985). *Der philosophische Diskurs der Moderne: Zwölf Vorlesungen*. Frankfurt a.M.: Suhrkamp.

Habermas, J. (1987). Excursus on Luhmann's Appropriation of the Philosophy of the Subject through Systems Theory. In *The Philosophical Discourse of Modernity: Twelve Lectures* (pp. 368-385). Cambridge, MA: MIT Press.

Hull, D. L. (2001). *Science and Selection: Essays on Biological Evolution and the Philosophy of Science*. Cambridge, UK: Cambridge University Press.

Husserl, E. (1929). *Cartesianische Meditationen und Pariser Vorträge. [Cartesian Meditations and the Paris Lectures.]*. The Hague: Martinus Nijhoff, 1973.

Husserl, E. (1936). Der Ursprung der Geometrie als intentional-historisches Problem. *Revue internationale de philosophie,* 1(2), 203-225.

Husserl, E. (1962). *Die Krisis der Europäischen Wissenschaften und die transzendentale Phänomenologie*. Den Haag: Martinus Nijhoff.

Huygens, C. (1888-1950). *Oeuvres Complètes*. The Hague: Nijhoff.

Künzler, J. (1987). Grundlagenprobleme der Theorie symbolisch generalisierter Kommunikationsmedien vei Niklas Luhmann. *Zeitschrift für Soziologie,* 16(5), 317-333.

Leibniz, G. W. (1695). New Systems of the Nature and of the Communication of Substances, and of the Union between the Soul and the Body, *Journal des Savants*, June 1695.

Leibniz, G. W. ($^3$1966). *Hauptschriften zur Grundlegung der Philosophie,* edited by E. Cassirer . Hamburg: Meiner.

Leibniz, G. W. ([1710] 1962). *Essais de Théodicée sur la Bonté de Dieu, la liberté de l'homme et l'origine du mal*. Paris: Aubier.

Levinas, E. (1953). Liberté et Comandement. *Revue de Métaphysique et de Morale,* 58, 264-272.

Levinas, E. (1963). La Trace de l'autre. *Tijdschrift voor Filosofie,* 25(1963), 605-623.

Leydesdorff, L. (1994). Uncertainty and the Communication of Time,. *Systems Research,* 11(4), 31-51.

Leydesdorff, L. (1995). *The Challenge of Scientometrics: The Development, Measurement, and Self-Organization of Scientific Communications*. Leiden: DSWO Press, Leiden University; at http://www.universal-publishers.com/book.php?method=ISBN&book=1581126816.

Leydesdorff, L. (2001). *A Sociological Theory of Communication: The Self-Organization of the Knowledge-Based Society*. Parkland, FL: Universal Publishers; at http://www.universal-publishers.com/book.php?method=ISBN&book=1581126956

Leydesdorff, L. (2003). The Construction and Globalization of the Knowledge Base in Inter-Human Communication Systems. *Canadian Journal of Communication,* 28(3), 267-289.

Leydesdorff, L. (2005). Anticipatory Systems and the Processing of Meaning: A Simulation Inspired by Luhmann's Theory of Social Systems. *Journal of Artificial Societies and Social Simulation,* 8(2), Paper 7, at http://jasss.soc.surrey.ac.uk/8/2/7.html.

Leydesdorff, L. (2006). "While a Storm is Raging on the Open Sea": Regional Development in a Knowledge-Based Economy. *Journal of Technology Transfer,* 31(1),189-203.

Leydesdorff, L. (forthcoming). Hyperincursion and the Globalization of a Knowledge-Based Economy. CASYS'05, 7th International Conference on Computing Anticipatory Systems. D. M. Dubois (Ed.), *AIP Conference Proceedings*, American Institute of Physics.

Leydesdorff, L., & D. M. Dubois. (2004). Anticipation in Social Systems: The Incursion and Communication of Meaning. *International Journal of Computing Anticipatory Systems,* 15, 203-216.

Leydesdorff, L., W. Dolfsma, & G. v. d. Panne. (2006). Measuring the Knowledge Base of an Economy in Terms of Triple-Helix Relations among 'Technology, Organization, and Territory'. *Research Policy,* 35(2), 181-199.

Luhmann, N. (1984). *Soziale Systeme. Grundriß einer allgemeinen Theorie*. Frankfurt a. M.: Suhrkamp.

Luhmann, N. (1986). The Autopoiesis of Social Systems. In F. Geyer & J. v. d. Zouwen (Eds.), *Sociocybernetic Paradoxes* (pp. 172-192). London: Sage.

Luhmann, N. (1990). *Die Wissenschaft der Gesellschaft*. Frankfurt a.M.: Suhrkamp.

Luhmann, N. (1995). *Die neuzeitlichen Wissenschaften und die Phänomenologie*. Vienna: Picus.

Luhmann, N. (1997). *Die Gesellschaft der Gesellschaft*. Frankfurt a.M.: Surhkamp.

Maturana, H. R. (2000). The Nature of the Laws of Nature. *Systems Research and Behavioural Science,* 17, 459-468.

Popper, K. R. (1972). *Objective Knowledge. An Evolutionary Approach*. Oxford: Oxford University Press.

Rosen, R. (1985). *Anticipatory Systems: Philosophical, Mathematical and Methodological Foundations*. Oxford, etc.: Pergamon Press.

Schutz, A. (1975). *Collected Papers III. Studies in Phenomenological Philosophy*. The Hague: Martinus Nijhoff.

Schütz, A. (1952). Das Problem der transzendentalen Intersubjectivität bei Husserl. *Philosophische Rundschau,* 5(2), 81-107.

Shannon, C. E. (1948). A Mathematical Theory of Communication. *Bell System Technical Journal,* 27, 379-423 and 623-356.

Theil, H. (1972). *Statistical Decomposition Analysis*. Amsterdam/ London: North-Holland.